\documentclass[10pt, conference, a4paper]{IEEEtran}

\usepackage{graphicx}
\usepackage{times}
\usepackage[latin1]{inputenc}
\usepackage{xspace}
\usepackage{url}
\usepackage{amssymb}

\newcommand{\ca}{\textit{Ca}\xspace}
\newcommand{\ce}{\textit{Ce}\xspace}

\begin{document}

\title{Legacy Software Restructuring: Analyzing a Concrete Case}

\author{Nicolas Anquetil, Jannik Laval}

\author{\IEEEauthorblockN{Nicolas Anquetil, Jannik Laval}
\IEEEauthorblockA{RMoD Team,\\
INRIA Lille-Nord-Europe\\
France\\
\emph{firstname.lastname}@inria.fr}
}

\maketitle

\begin{abstract}
Software re-modularization is an old preoccupation of reverse engineering research.
The advantages of a well structured or modularized system are well known.
Yet after so much time and efforts, the field seems unable to come up with solutions that make a clear difference in practice.
Recently, some researchers started to question whether some basic assumptions of the field were not overrated.
The main one consists in evaluating the high-cohesion/low-coupling dogma with metrics of unknown relevance.
In this paper, we study a real structuring case (on the Eclipse platform) to try to better understand if (some) existing metrics would have helped the software engineers in the task.
Results show that the cohesion and coupling metrics used in the experiment did not behave as expected and would probably not have helped the maintainers reach there goal.
We also measured another possible restructuring which is to decrease the number of cyclic dependencies between modules.
Again, the results did not meet expectations.
\end{abstract}

\begin{IEEEkeywords}
Re-modularization, re-structuring, cohesion, coupling, metrics, case study
\end{IEEEkeywords}

\section{Introduction}

Restructuring old software systems aims at breathing new life into them and allow to maintain them more easily for a longer period of time.
For years, research tried to support this activity by proposing ways to improve the modularity of the system as measured by different cohesion and coupling metrics.
The dogma is that good modularization should exhibit high cohesion and low coupling (e.g. \cite{Anqu03a,Hall05a}).
Cohesion and coupling metrics were measured by a variety of metrics but that all tend to rely on syntactical aspects of the source code (with some exceptions).

It must be recognized that after more than a decade of research, very little of this work seems to have made its way to the industry.
Recently, some voices started to raise doubts on the cohesion/coupling dogma in some way or another \cite{Abre01a,Bhat06a,Sind07a}.

We propose that one of the problems of the cited research is that they studied re-structuring techniques and/or metrics without first firmly established whether the metrics ---or even the cohesion/coupling dogma--- had any relevance.
Results were evaluated using the same (or similar) metrics used to obtained them (see \cite{Anqu99a, Anqu03a}) and against decompositions of the systems of unknown value (typically the actual structure of the systems used).

In this paper, we propose an experiment setup to test the relevance of cohesion/coupling metrics alone, without considering the results they might propose when used with one or the other restructuring algorithm.
The metrics are evaluated against their expected results on two successive versions of a real life system (the Eclipse platform) that went through an explicit restructuring effort.

In the following sections, we will present past research on software remodularization and cohesion/coupling metrics, we also set our working hypothesis in contrast to the previous research (Section \ref{sec:cohcop}); In Section \ref{sec:eclipse}, we present our case study, two successive restructurings of the Eclipse platform; we then (Section \ref{sec:metrics}) present the modularization quality metrics we put to the test; Section \ref{sec:results} discusses the results of the experiments; and we conclude in Section \ref{sec:conclu}.

\section{Software Remodularization}
\label{sec:cohcop} 

Research in the field usually talk about the quality of modularization and one could, therefore, talk of re-modularization approaches.

We will generally prefer the term of re-structuring which is more generic, the goal of a restructuring must be to change the structure of a system for some defined goal, for re-modularization, the goal implicitly is to get a more modular system.
Considering the lack of resources for software maintenance, it is very likely that re-structuring does happen more often than strict re-modularization in real life.
That is to say, organizations need a stronger incentive than ``mere'' modularization optimization to undertake such large and radical task as a full system re-structuring.
This simple difference in terms might mean that new quality metrics will have to be proposed to take into account new dimensions of restructuring apart from improving the modularity quality.
This possibility will not be further discussed in this paper.

\subsection{Past Work on Re-structuring and Cohesion/Coupling}

Good modularization count as one of the few fundamental rules of good programming.
According to this rule, systems must be decomposed in modules (or packages, subsystems, namespaces, etc.) that have high cohesion and low coupling.
The heralded advantages of a modular architecture includes \cite{Bhat06a}:  handle complexity of a large system; design and develop different parts of same system by different people; test system in partial fashion; repair defective parts of a system without interfacing with other parts; control defect propagation; or, reuse existing parts in different contexts.
Some coupling metrics have been found to be good predictors of fault proneness (e.g. \cite{Bink98a,Bria97a}),
and a model including coupling metrics was shown to be a good predictor of maintenance effort \cite{Li93a}.

The high cohesion, low coupling rule may be interpreted in various ways \cite{Sind07a}.
For example, semantically, high cohesion would mean that all the components of a module share the same purpose (that of the module), termed singularity and similarity of purpose in \cite{Sind07a}.
Low coupling would mean that this purpose is not shared by components of other modules (or to a lesser degree).
Because computers are not good at dealing with semantics, other interpretations, easier to measure, are usually preferred, for example based functional dependencies ---one component calls a function of another component--- (e.g. \cite{Abre01a,Anqu99a,Anqu03a,Bhat06a}), or on data access (e.g. \cite{Dave00a,Hall05a}), or co-changes in a version control system (e.g. \cite{Beye05b,Gall03a}).
In \cite{Bria98a}, Briand, Daly and W\"ust identified and organized more than thirty coupling metrics.

However, these metrics seem to have their limits and some researchers started to question the utility of cohesion and coupling as design quality criteria: \cite{Abre01a,Bhat06a,Sind07a}: 
Even Briand \textit{et al.} remark that ``the usefulness of many measures has yet to be demonstrated'' \cite{Bria98a}.

Thus, there are proposition to include new metrics, such as the size of the modules (a module should be neither too small, nor too big) \cite{Abre01a,Anqu99a,Anqu03a,Bhat06a} to introduce new points of view on the problem.

Note that there is also a large body of work on the clustering algorithm(s) susceptible to produce the best modularization (e.g. \cite{Anqu99a,Anqu03a,Beye05b,Bhat06a,Dave00a,Manc99a}).
The present research must be seen as a preliminary, to find a good algorithm one needs first a proper quality criterion.
Since, there is, as yet, no clear proof that the existing modularity quality metrics are of any use, there is still a possibility that new metrics, not suited for some clustering algorithms, will need to be found.

We see a fundamental problem in all this research: the difficulty to validate empirically the validity of the quality metrics (a fact alluded to in \cite{Abre01a,Anqu99a,Anqu03a,Beye05b} for example).

Considering that ``science rides on the wings of its measuring instruments'', we decided to look for a new evaluation methodology for modularization quality metrics.

\subsection{Evaluation of the Modularization Quality Metrics}
\label{sec:hypotheses}

Abreu and Goul\~ao's paper \cite{Abre01a} appears to question the validity of cohesion and coupling as design quality criteria.
Yet, they measure the modularization improvement of their solution in terms of how high the cohesion and low the coupling are.
Along with them, others (e.g. \cite{Anqu99a,Anqu03a}) uses the current decomposition of a system to assess the quality of their proposed modularization.
They are, therefore, in the uncomfortable situation of evaluating the adequacy of quality metrics they propose, using the same quality metrics to measure their progress in comparison to a modularization of unknown value.
The problem is made even worse by the recognized fact that for one system there are multiple possible modularizations that are probably equally valid, in short ``different'' does not mean better (or worst).


We believe that research has been trying to solve too many problems too soon and decided to go back to the base hypothesis of modularization quality metrics.
That is to say try to evaluate whether the existing metrics are useful, and if not, what other metrics are needed.
We need to tackle two issues: (i) we need a modularization of known value to be able to assess the relevance of the metrics; and, (ii) we need an evaluation method that will take into account the fact that different points of view on modularization exist and therefore different modularizations may be equally valid.

To tackle issue (i), we hypothesize that a software system just after a remodularization effort by its maintenance team could be considered to have a good modularization.
The hypothesis is obviously subject to discussion, but it is credible if we focus on past remodularizations of systems that stood the proof of time.
We may actually lessen this hypothesis by saying that the modular quality of a system should improve after an explicit restructuring effort by its maintenance team.
A similar hypothesis was informally used by Sarkar \textit{et al.} in \cite{Sark07a}.
We therefore propose to evaluate the relevance of quality metrics by comparing their results in two successive versions of system that went through an explicit restructuring effort.

\begin{figure*}[!t]
  \begin{center}
    \includegraphics[width=0.45\textwidth]{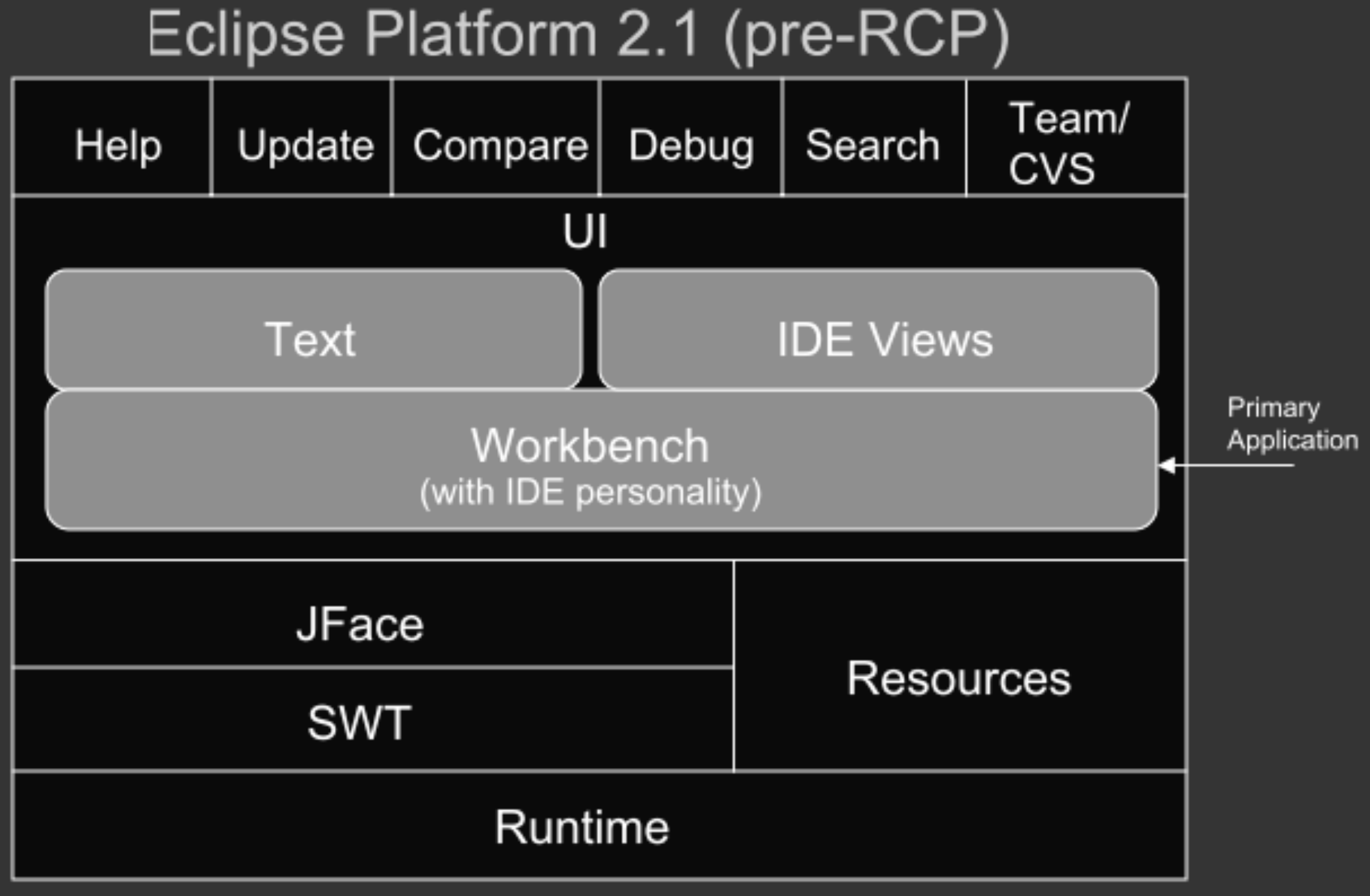}
    \hspace{1em}
    \includegraphics[width=0.473\textwidth]{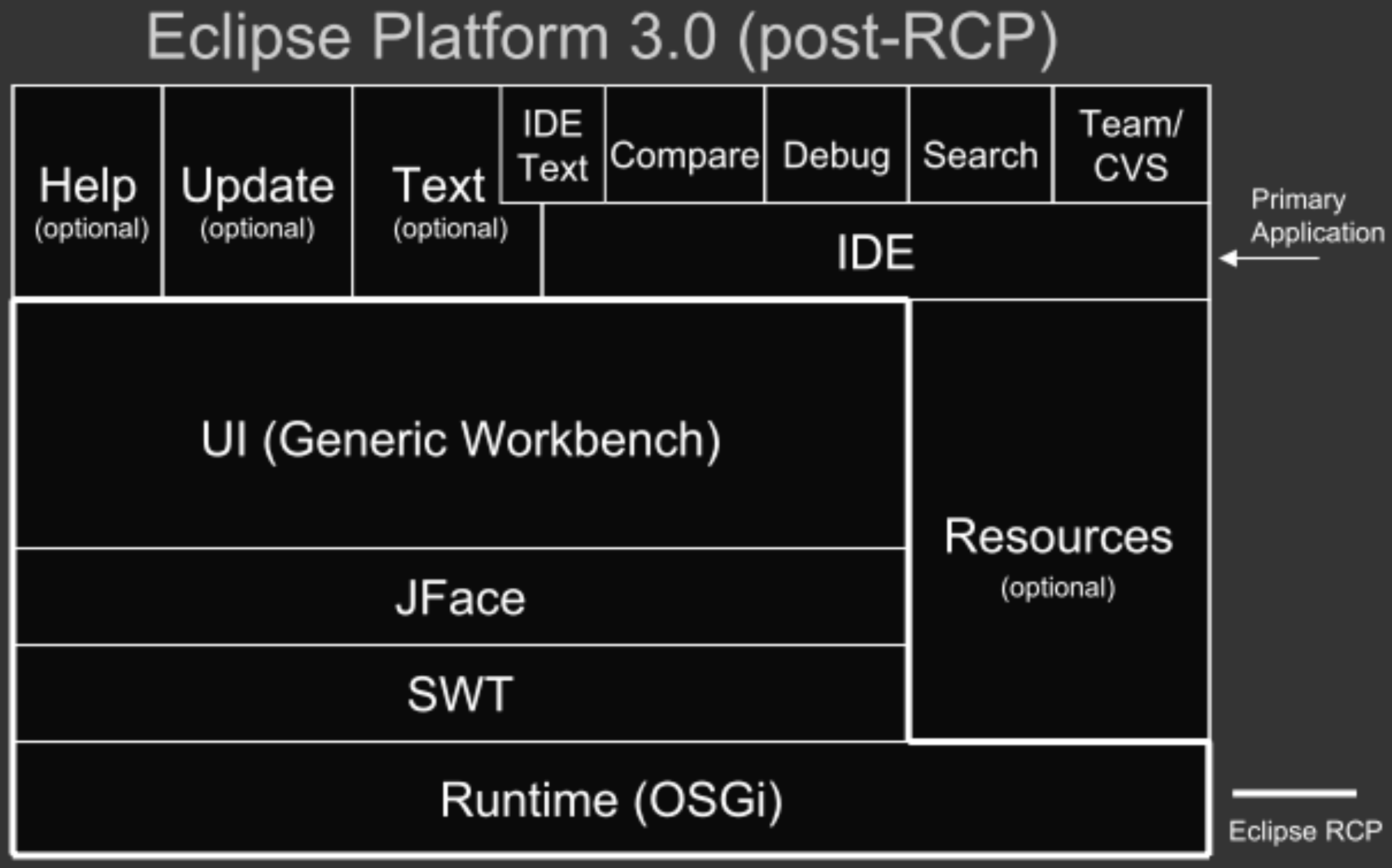} 
    \caption{An illustration of the architecture of the Eclipse platform before RCP (v. 2.1) and after RCP (v. 3.0). From a presentation at EclipseCon 2004.}
    \label{fig:eclipsecon}
  \end{center}
\end{figure*}

The fact that we compare two modularizations of the same system also tackles issue (ii).
Because we are not evaluating one isolated modularization, but comparing two modularizations of know values (relative one to the other), we don't need to consider that different modularizations may be equally valid.
In our case, we are assuming that the second should be better than the first.
Again, this hypothesis is subject to discussion, for example, one could argue that the earlier version of the system was structured according to one point of view and the later was structured according to another.
In this sense, the two structures would be different but not clearly better or worst.
If this situation is theoretically valid, it is practically very unlikely:
\begin{itemize}
\item The point of views between the two structures cannot be very much different, otherwise a ``simple'' system restructuring would not be enough to attend the new needs.
One restructures when the system is almost good enough, otherwise a new system or at least of significant part of the new system needs to be created.
If the two views are not that much apart one from the other, the notion of two different points of view with equal value becomes less likely;

\item One may assume that professionals would not dedicate their time to an explicit restructuring effort and not try to achieve the best result possible.

\item For more security, one can limit oneself to restructuring of mature systems that already had a long maintenance history behind them.
Considering Lehman's law of software evolution \cite{Lehm96a}, such systems would already have suffered a decrease in their quality due to the maintenance.
This would increase the probability that the restructuring has a better modular quality.

\item Finally, for even more security, one should repeat the tests on a number of restructuring cases to ensure that the results do not depend on a unique, possibly not adequate case study.
\end{itemize}

\section{The case study: Eclipse RCP}
\label{sec:eclipse} 

The focus of our study was the restructuring that occurred in the Eclipse platform between versions 2.1 and 3.0, when eclipse went from the concept of an extensible IDE (v. 2.1) toward a ``Rich Client Platform'' (v. 3.0, Eclipse RCP).
A graphical representation of the old and new structure of Eclipse is pictured in Figure \ref{fig:eclipsecon} (from a presentation at EclipseCon 2004\footnote{\url{http://www.eclipsecon.org/2004/EclipseCon_2004_TechnicalTrackPresentations/11_Edgar.pdf}}).
We also studied the precedent version (v. 2.0.1) where some preliminary restructuring took place; and the following version (v. 3.1) to check whether some issues identified just after restructuring were mended in the following version (maturation of the structure).
In total we studied four successive versions of the system that represent three evolutions.
Each evolution will be denoted by the version numbers it comes from and goes to, e.g. 2.1$\rightarrow$3.0 .

In reference to the validity of our hypotheses (see \ref{sec:hypotheses}), we may cite this line from \url{http://eclipse.org/rcp/generic_workbench_structure.html}: ``Prior to 2.1, the org.eclipse.ui plug-in was the monolithic implementation of the Eclipse Platform UI. The above picture reflects the restructuring that done for 2.1 which introduced several new plug-ins''.
This reference to ``monolithic implementation'' clearly denotes an acknowledgement of a modularization issue, we can only suppose that the developers did try to solve this issue by introducing the new structure.

For the sake of discussion, we will call the two first evolutions (2.0.1$\rightarrow$2.1 and 2.1$\rightarrow$3.0) ``restructurings'' in the sense that they explicitly purpose to restructure the system in some way (even if this is not their sole goal).
On the contrary, the last evolution (3.0$\rightarrow$3.1) will be simply called an evolution in the sense that it exhibits no explicit restructuring purpose.

Documents describing the restructurings were found at: \url{http://wiki.eclipse.org/index.php/Rich_Client_Platform}

We used a derivative of the Java Development Toolkit (JDT) parser to extract information from the source code.

Eclipse is a multi-platform environment and includes code (particularly for the GUI) for different window environments: Windows, Linux, Mac.
We elected the GTK (Linux) version for our experiments.

For version 2.0.1, we tried to select the minimum set of eclipse plugins needed to build the environment.
For the successive versions, we followed the same rule, based on the plugins already selected in the preceding version so as to ensure a maximum of continuity in the software components included in each version.

\section{The metrics used in the study}
\label{sec:metrics} 

\subsection{Descriptive statistics}
\label{sec:met-size} 

We will use different size metrics to have a basic understanding of what happened between two successive versions of the case study.

These metrics are: Number of Java packages, number of Eclipse plugins, number of classes, number of methods, number of lines of code, number of method invocations.

These metrics will give us an approximative understanding of the size of the system (number of Java packages, Eclipse plugins, classes, methods, and lines of code) and the ``density'' of interaction inside the system (number of method invocations).

\begin{table*}[!t]
\begin{center}
\caption{Descriptive statistics on four successive versions of the Eclipse platform}
\label{tab:res-size}
\begin{tabular}{ccccccc}
\hline
vers. & \# packages & \# plugins & \# classes & \# methods & LOC & \# invocations \\ 
\hline
2.0.1	& 101 & 10 & 3\,209 &~23\,172 & 417\,109 & ~53\,302 \\ 
2.1		& 144 & 18 & 4\,034 & 29\,098 & 540\,948 & ~66\,806 \\ 
3.0		& 251 & 26 & 6\,449 & 44\,377 & 804\,071 & 100\,667 \\ 
3.1		& 307 & 26 & 7\,612 & 52\,369 & 969\,078 & 115\,541 \\ 
\hline
\end{tabular} 
\end{center}
\end{table*}

\subsection{Cohesion/Coupling metrics}
\label{sec:met-coco} 

Our main purpose is to evaluate whether existing cohesion/coupling metrics are a good measure of the quality of a modularization.

In this paper we present the results for the following cohesion/coupling metrics: Bunch \cite{Manc99a} cohesion/coupling metrics, \ca and \ce \cite{Mart94c}.

We compute Bunch coupling between modules (i.e. Java packages and Eclipse plugins, see later) as follows:
\begin{itemize}
\item We say that a class $c_1$ depends on a class $c_2$ if we can find at least one method invocation from a method of $c_1$ to a method of $c_2$\footnote{Because of dynamic dispatch, it is impossible to specify clearly which class $c_2$ is considered ``owner'' of the invoked method, it can be the class that \textit{declares} the method or the class that \textit{implements} it or one of their \textit{descendants} (see \cite{Bria98a}).
We considered the class that the JDT binding resolution mechanism indicates as the owner of the invoked method.}.
This is \textit{import coupling} in Briand's terminology \cite{Bria98a}.

\item Cohesion ($A$) of a module $i$ with $N_i$ classes is the number $\mu_i$ of intra dependence edges between classes of the module, normalized by the maximum number of possible dependences:
\[A_i=\mu_i / N^2_i\]

\item Coupling ($E$) between a module $i$ (with $N_i$ classes) and a module $j$ (with ($N_j$ classes) is the number ($\epsilon_{i,j}$) of dependence edges from a class in $i$ to a class in $j$, normalized by the maximum number of possible dependences:
\[E_{i,j}=\epsilon_{i,j} / 2N_iN_j\]

\item Coupling of a module $i$ is the sum of its coupling to all the other modules of the system:
\[E_i=\sum_{j\neq i}A_{i,j}\]
\end{itemize}

We also use method invocation to compute \ca and \ce:
\begin{itemize}
\item We extend the dependence relationship from classes to their containing module (Java package or Eclipse plugin): a module $i$ depends on a module $j$ if at least one class of $i$ depends on one class of $j$;

\item $Ca_i$ is computed as the number of classes outside module $i$ depending on (some classes inside) module $i$. Munnelly \cite{Munn07a} states that ``a high value is an indicator of responsibility''.

\item $Ce_i$ is computed as the number of number of classes outside module $i$ on which (some classes inside) module $i$ depends. Munnelly \cite{Munn07a} states that ``the lower the value, the more independent the module''. We will be more interested in \ce than in \ca.
\end{itemize}

These metrics were computed for Java packages and Eclipse plugins.
Java packages are the usual modularization scheme in Java that allows to separate classes in individual modules.
To use a class (or one of its member) from another package than its own, a Java class must import it.
On the other hand,  by default\footnote{Access right: ``default package''} a class has access to all the other classes in its own package and their members.
We studied the cohesion/coupling of Java packages because they are the ``natural'' modularization structure for Java systems.
Eclipse plugins or a deployment scheme that allows to package some functionality in a `.jar' file and distribute it easily.
Eclipse plugins are a coarser grained decomposition of the system, they usually contain more than one Java package.
Although they may not be frequently used that way, they are also a perpendicular decomposition of the system: a Java package may be scattered over two or more Eclipse plugins.
This does happen in our case study.
We studied the cohesion/coupling of Eclipse plugins because they are deployment entities used in the eclipse ecosystem.
As such, one may hypothesize that developers would pay closer attention to their coupling so that they can be deployed individually.

\subsection{Cyclic Dependencies}
\label{sec:met-cycle} 

Cyclic dependencies are known to introduce extra difficulty in the maintenance process as all packages in the a cycle will depend on all other packages in the same cycle (see the ADP ---Acyclic Dependencies Principle--- in \cite{Mart00b}).
For example, it means that packages can no longer be reused independently, one must take the whole cycle; or
ripple effects must always be computed on all packages of the cycle.
Apart from improving the modularity of the system, re-structuring could also aim at removing cyclic dependencies between packages (e.g. see \cite{Lava09c,Sark07a}).

We computed the number and size of Strongly Connected Components (SCC) in the system as an indicator of the presence and importance of cyclic dependencies between Java packages or Eclipse plugins.

An SCC in a graph is a maximal set of vertices (i.e.,  modules)  in  which  there  exists  a  path  from  every
vertex in the SCC to every other vertex in the same set.
An SCC may contain more than one cycle if one vertex belongs to more than one cycle.
So the number of SCCs is not an exact measure of the number of cyclic dependencies in the system, but it can serve as an indicator.
To complement this indicator, we also monitored the size (in number of vertices) of the largest SCC.

\section{Results}
\label{sec:results} 

\begin{table*}[!t]
\begin{center}
\caption{Comparing the evolution (increase, decrease, or stationary) of the cohesion and coupling (Bunch metrics \cite{Manc99a}) of Java packages and Eclipse plugins in four successive versions of the Eclipse platform}
\label{tab:local-bunch}
\begin{tabular}{r@{$\rightarrow$}lc@{ }c@{ }cc@{ }c@{ }ccc@{ }c@{ }cc@{ }c@{ }c}
\hline
\multicolumn{2}{c}{}&\multicolumn{6}{c}{JAVA PACKAGES} && \multicolumn{6}{c}{ECLIPSE PLUGINS} \\
\multicolumn{2}{c}{} & \multicolumn{3}{c}{Cohesion} & \multicolumn{3}{c}{Coupling} && \multicolumn{3}{c}{Cohesion} & \multicolumn{3}{c}{Coupling}\\
\multicolumn{2}{c}{} & incr. & same & decr. & incr. & same & decr.	&& incr. & same & decr. & incr. & same & decr. \\
\cline{3-8} \cline{10-15}
2.0.1&2.1	& 12 & 34 & 44 & ~23 & 12 & 59		&& 2 & 0 & ~7 & 5 & 2 & ~2 \\ 
2.1&3.0		& 32 & 49 & 58 & ~48 & 21 & 70		&& 3 & 0 & 13 & 0 & 3 & 14 \\ 
3.0&3.1		& 64 & 78 & 98 & 115 & 28 & 97		&& 1 & 1 & 23 & 4 & 3 & 18 \\ 
\hline
\end{tabular} 
\end{center}
\end{table*}

\subsection{System size}
\label{sec:res-size} 

Table \ref{tab:res-size} gives some descriptive statistics on the four versions studied: the number of Java packages, the number of Eclipse plugins, the number of classes and methods, the number of lines of code (LOC) and the number of method invocations found.

One can see that there is a constant increase in all the items (save the number of plugins which remains constant between versions 3.0 and 3.1) which correspond to the accepted opinion that software systems tend to get more complex and to grow as they get older (2nd and 6th law of software evolution \cite{Lehm96a}).

\subsection{Bunch Cohesion/Coupling Results}
\label{sec:res-bunch} 

According to the good modularization principles, a good re-structuring of the system should increase its cohesion and decrease its coupling.
Automatic re-structuring techniques try to achieve this by optimizing ``local'' cohesion/coupling of every module.
It would also seem acceptable to try to optimize ``globaly'' cohesion/coupling while accepting some local degradation.

To check whether the restructurings that took place between versions 2.0.1 and 2.1 and between versions 2.1 and 3.0 improved cohesion/coupling, we will focus on the increases and decreases of the metrics between two successive versions of the same item, a Java package or an Eclipse plugin.
Note that all packages (or plugins) may not exist in all versions, new ones may be created, and old ones may be removed.

Table \ref{tab:local-bunch} gives the number of Java packages (on the left) for which cohesion (resp. coupling) increased, stagnated or decreased between two successive versions.
We are more interested in knowing how many packages increased their cohesion (should be high) vs. the number that decreased it (should be low).
For coupling, the expectation is the opposite, more packages should decrease their coupling than packages than increase it.

We can see that, for the two restructurings (2.0.1$\rightarrow$2.1 and 2.1$\rightarrow$3.0), the coupling does tend to decrease for more packages than it increases.
For the third evolution (3.0$\rightarrow$3.1), more packages increase their coupling than decrease it.
Overall, the two restructurings seemed to, somehow, improve locally the coupling of more packages than degraded it.
Globally, the average coupling of the packages in the system (see Table \ref{tab:global-bunch}) does not point toward a clear improvement, the first restructuring increased the global coupling (degradation), the second results in a small improvement, and the evolution shows a stagnating coupling.

\begin{table}[!t]
\begin{center}
\caption{Average Cohesion and coupling (Bunch metric) for four successive versions of Eclipse}
\label{tab:global-bunch}
\begin{tabular}{cccccc}
\hline
 & \multicolumn{2}{c}{JAVA PACKAGES}  && \multicolumn{2}{c}{ECLIPSE PLUGINS}  \\ 
 & Cohesion & Coupling && Cohesion & Coupling \\ 
 \cline{2-3} \cline{5-6}
2.0.1 & 0.099 & 0.058 && 0.036 & 0.004 \\ 
2.1   & 0.087 & 0.068 && 0.033 & 0.005 \\ 
3.0   & 0.081 & 0.065 && 0.034 & 0.004 \\ 
3.1   & 0.076 & 0.065 && 0.025 & 0.003 \\ 
\hline
\end{tabular} 
\end{center}
\end{table}

On the other hand, when one turns to cohesion, the results seem contradictory.
Whereas one usually expect cohesion to have opposite trend than coupling (decreasing coupling is associated with increasing cohesion), we see here the same trend, i.e. locally (Table \ref{tab:local-bunch}), more packages decreased their cohesion than increased it, and globally (Table \ref{tab:global-bunch}), the average cohesion of the system is decreasing with each new version.

For the Eclipse plugins, the results are similar.
As exhibited in Tables \ref{tab:local-bunch} (right) and \ref{tab:global-bunch}.
The coupling decreases for more plugins than increases (except for the restructuring 2.1$\rightarrow$3.0 where only 2 packages decreased coupling and 5 increased it), which seems a positive trend.
The coupling, however, also decreases for more plugins than the opposite.

Globally, the average cohesion and coupling results do not exhibit a simple trend.
The numbers don't tell a straight story, but they clearly do not advocate for an improved cohesion/coupling over successive versions: global coupling may be considered either stationary or slightly decreasing; and, apart for a small increase in the second re-structuring (2.1$\rightarrow$3.0), cohesion is decreasing.
The only positive point from the cohesion/coupling point of view is that coupling is much lower than cohesion for Eclipse plugins, a fact that we did not see, so clearly, in the Java packages.


To summarize, we end up with several findings that do not agree with the general understanding of cohesion and coupling:
\begin{itemize}
\item The restructuring did not exhibit modularity improvement as measured by the Bunch cohesion/coupling metrics (i.e. they don't show cohesion increase and coupling decrease);

\item More often than not, cohesion and coupling present the same evolution (a decrease) instead of opposite evolution;

\item Their does not seem to be any clear difference in trends between restructurings (2.0.1$\rightarrow$2.1 and 2.1$\rightarrow$3.0) and evolution (3.0$\rightarrow$3.1).
\end{itemize}

\begin{table*}[!t]
\begin{center}
\caption{Comparing the evolution (increase, decrease, or stationary) of afferent (\ca) and efferent (\ce) coupling of Java packages and Eclipse plugins in four successive versions of the Eclipse platform}
\label{tab:res-cace}
\begin{tabular}{r@{$\rightarrow$}lc@{ }c@{ }cc@{ }c@{ }ccc@{ }c@{ }cc@{ }c@{ }c}
\hline
\multicolumn{2}{c}{}&\multicolumn{6}{c}{JAVA PACKAGES} && \multicolumn{6}{c}{ECLIPSE PLUGINS} \\
\multicolumn{2}{c}{} & \multicolumn{3}{c}{\ce} & \multicolumn{3}{c}{\ca} && \multicolumn{3}{c}{\ce} & \multicolumn{3}{c}{\ca}\\
\multicolumn{2}{c}{} & incr. & same & decr. & incr. & same & decr.	&& incr. & same & decr. & incr. & same & decr. \\
\cline{3-8} \cline{10-15}
2.0.1&2.1	& ~52 & 33 & 13 & ~58 & 26 & 14	&& ~5 & 2 & 2 & ~6 & 2 & 1 \\ 
2.1&3.0		& ~75 & 43 & 25 & ~88 & 38 & 17	&& 12 & 2 & 3 & 10 & 4 & 3 \\ 
3.0&3.1		& 119 & 72 & 53 & 124 & 79 & 41	&& 15 & 6 & 4 & 18 & 4 & 3 \\ 
\hline
\end{tabular} 
\end{center}
\end{table*}

Some explanations may be proposed.

The fact that the restructurings did not show improvement in the cohesion/coupling metrics, might be due to the case study itself.
We plan to investigate more systems in the future to see if they behave the same way.
Yet, it is little likely that this particular restructuring should not be considered successful: six years latter, Eclipse is still alive and well, and the RCP architecture that was introduced at that time is still in use.
There is, therefore, a real possibility that we should reconsider the expectations we have on a good restructuring, and in particular in the usual belief\footnote{One of the conference reviewer objected to our frequent use of the term ``belief'', arguing that science should not be based on beliefs. Our point is exactly that high cohesion/low coupling is accepted without real proof of its value, a value that we are challenging here.} that restructuring should improve the modularity of system as measured by the Bunch metrics.

If we accept that, we can reconsider the whole modularity improvement belief, or we can blame the Bunch metrics we used, i.e. we can still think that restructuring should improve the modularity of the system, but that the Bunch metrics used here do not measure modularity or only measure a partial aspect of modularity.

Actually the Bunch metrics we used, do present a problem that causes the decrease trend of both metrics whereas the usual understanding is that they evolve in opposite directions.
In summary, the metrics must both decreases because their numerator show a linear increase when their denominator show a quadratic increase (see also Cohesion and Coupling formula, Section \ref{sec:met-coco}, and data in Table \ref{tab:res-size}): (i) the number of classes steadily increases between versions; (2) the number of methods invocations also increases, linearly to the number of classes; (3) the Bunch cohesion/coupling metrics are normalized by the maximum number of possible pairs (of classes) which show a quadratic increases with respect to the number of classes.
We will look for other cohesion/coupling metrics that would not, a priori, present this issue.

The lack of difference between restructuring and evolution could be due to the fact that, from a cohesion/coupling point of view, they naturally present no differences. It could also be a specificity of this particular case study for which restructurings were not only pure restucturings, but also introduced new features and corrected bugs.
This issue should be tested by experimenting with other system restructuring efforts.

\subsection{Afferent/Efferent Coupling Results}
\label{sec:res-cace} 

As explained in Section \ref{sec:met-coco}, Afferent coupling (\ca) is not really a measure of coupling in the sense we use it here, but rather an indicator of responsibility of the module.
Yet for the good modularity of the system, (global) \ca should be low.
Efferent coupling (\ce)is a coupling metrics in the same sense of the Bunch coupling metric, i.e. it is an indicator of module independence (or lack of).
For the good modularity of a package and of the system, local and global \ce should be low.
\ce also presents the advantage over Bunch coupling that it does not include quadratic variation in its formula because it just counts the number of external classes on which a module depends on (see discussion at the end of previous section).

Results for this experiment are given in Table \ref{tab:res-cace}.
We are primarily interested in the results of the \ce metric.

For Java packages, one can see that the majority of packages present an increasing \ce.
If we consider increasing and stationary \ce, we get about 80\% of the Java packages.
This is not the expected result of a restructuring that should decrease coupling.
\ca shows the same trend, but that would be expected since the two metrics are two faces of the same quantity.

For Eclipse plugins, the results of \ce and \ca are similar with about 80\% of the plugins with either increasing or stationary \ce.

This new experiment seems to point toward the same conclusions as the previous one with the Bunch metrics: contrary to the expected, the system restructuring did not improve the coupling of the system as measured by the \ce metric.
We have no cohesion metric here to see whether it would have evolved similarly to \ce.
And there does not seem to be any significant differences between the two restructurings and the simple evolution.

The possible explanations for this would be the same as in the previous experiment, with the difference that the coupling metric does not mathematically produce a diminution of the coupling with the increase in number of classes, methods or invocations.
And actually, using \ce, that removes the issue we detected in the Bunch cohesion/coupling metrics, we do see an increase in the coupling that follows the increase in class, method or invocation numbers.

\subsection{Cycles Results}
\label{sec:res-cycle} 

As part of this experiment, we were also interested in measuring the quality of the system in terms of cyclic dependencies between modules.
Cyclic dependencies are denounced as a bad coding practice for example in \cite{Mart00b}.
Research exists that tries to break such cycles to improve the modularity of systems  (e.g. \cite{Lava09c,Sark07a}).

\begin{table}[!t]
\begin{center}
\caption{Number of Strongly Connected Components (SCC) and largest SCC in the Java packages of four successive versions of the Eclipse platform}
\label{tab:scc}
\begin{tabular}{ccc}
\hline
version & \# SCC & Largest SCC \\ 
\hline
2.0.1 & 12 & 16 \\ 
2.1   & 13 & 21 \\ 
3.0   & 22 & 48 \\ 
3.1   & 23 & 66 \\ 
\hline
\end{tabular} 
\end{center}
\end{table}

We used two metrics to evaluate the system quality in terms of existence (or absence) of cycles: number of Strongly Connected Components (SCC) in the system and size of the largest SCC.
As explained in \ref{sec:met-cycle}, keep in mind that one SCC may involve more than one cycle if one vertex is part of two, or more, cycles.
Therefore, theoretically, there could be no relation between the number of SCC and the number of cycles.
In practice, it seems unlikely and the metrics was already used by others as in \cite{Sark07a}.

The results for Java packages are presented in Table \ref{tab:scc}, they show a clearly degrading situation where, with each new version, the number of SCC in the system increases.
The size of the largest SCC also increases, and at a faster pace.
This would indicate a parallel increase in the number and size of the cyclic dependencies between Java packages.

The two restructurings performed badly for these metrics.
This is one of the reason that prompted us to also consider version 3.1, to check whether the presence of cycles could not be seen as a temporary evil, accepted to allow finishing the restructuring into the desired RCP architecture, and that would have been removed later, after the new architecture had stabilized.
The results clearly show that this is not the case as the number and size of SCC continue to increase.

We don't give results for Eclipse plugins because we found no SCC, hence no cyclic dependencies among them.
Eclipse plugin did perform better in this sense.
It must first be noted that Eclipse forces the programmers to explicitly declare dependencies between plugins to be able to use the software components of one plugin into the other (note: we did not use this mechanism here but computed dependencies from the method invocations as for Java packages).
Eclipse also prohibit defining cyclic dependencies this way.
So the tool itself forced the engineers to pay attention to this issue.

It must also be noted that there are much less plugins (than Java packages) which greatly simplifies the task of checking for cyclic dependencies.

A third possible explanation for the abscence of cyclic dependencies between the Eclipse plugins, could be that these plugins are the ``real'' modules that the engineers considered (as opposed to the Java packages that would serve other purposes), and that as such, they paid more attention to their good modularity.
This explanation seems coherent with the description given in the Eclipse RCP restructuring documentation.
On the other hand, the two preceding experiments do not agree with this explanation and we could not identify any better modularity in terms of cohesion/coupling measured for Eclipse plugins.
With only one case study, we cannot offer more conclusive results.

We summarized and discuss our findings in the next section.

\section{Discussion and Conclusion}
\label{sec:conclu}

In these experiments, we studied the evolution of different metrics that purport to evaluate the quality of a modularization.

The traditional view of the problem is that restructuring a system should increase the cohesion of the modules, and decrease their coupling.
Another valid goal for a restructuring is to remove cyclic dependencies between modules which would tend to also reduce their coupling (cyclic dependencies are removed by breaking some dependency, thus reducing the coupling of at least one module.

We measured coupling between Java packages and Eclipse plugins with two different coupling metrics: Bunch's cohesion/coupling and Afferent/Efferent coupling (resp. \ca/\ce).
All metrics gave results that did not match expectations: After each of two restructurings, the Bunch cohesion did not increase and rather decreased, and efferent coupling, \ce, increased.

If we accept that the restructurings studied were reasonably successful ---which is a plausible hypothesis given the continuing success of the Eclipse platform six years later--- we remain with various possible explanations:
\begin{itemize}
\item The metrics themselves are flawed, or they do not adequately measure the modularity of systems;
\item The cohesion/coupling dogma is flawed, restructuring does not generally improve modularity in terms of cohesion/coupling;
\item There may be different types of restructurings that would have different properties, and here we would be in the presence of restructuring that did not improve the modularity of the system.
\end{itemize}

We already showed that the decreased in Bunch's cohesion and coupling was caused mechanically by a property of the metrics' formula rather than by a property of the system's modules themselves.
Therefore these particular metrics did present some kind of flaw.
However that does not apply for \ce that did not either give the expected results.
Because \ce is a very simple metric, it is hard to imagine that it has some other hidden flaw.
Other measurement with other coupling metrics could still be conducted to bring more light on this issue.

A second unexpected result was that cohesion and coupling (as measured with the Bunch metrics) do not evolve in opposition (one increasing when the other decreases), but presented the same trend (decreasing in our experiments) 

The third unexpected result, was that restrucurings did not either improved the situation in regard to the number of cyclic dependencies between the Java packages.

And, finally, a fourth unexpected results was that we could find no clear differences between restructuring evolutions and ``normal'' ones.

We plan to perform the same study on more systems to see whether they also lead to the same conclusions.
This would allow us to ascertain whether there are actually different kinds of restructuring activity, and perhaps to categorize these kinds.
The current high cohesion/low coupling dogma does not argue for this possibility.
But if it was lessened by further experiments similar to this one, identifying different kinds of restructuring would become a valid research path.

For now, we favor the hypothesis that the cohesion/coupling dogma, \textbf{as measured by the existing cohesion/coupling metrics}, is largely over-rated.
This does not automatically  mean one should not zeal for high cohesion/low coupling when developing or restructuring systems, but rather that we don't have any automated metrics that can measure the kind of cohesion and coupling the software engineers try to optimize in their development and restructuring work.
One must consider that software engineers work with higher level concepts that cannot be measured by the simple, existing, cohesion/coupling metrics.

\bibliographystyle{IEEEtran}
\bibliography{rmod,others}
 
\end{document}